\documentclass[lettersize, journal]{IEEEtran}
\usepackage{float}
\usepackage{caption}
\usepackage{color}

\usepackage{cite}
\usepackage{amsmath,amssymb,amsfonts}
\usepackage{amsthm}
\usepackage{graphicx}
\usepackage{textcomp}
\usepackage{xcolor}
\usepackage{svg}
\usepackage{subfigure}
\usepackage{mathtools}

\usepackage[ruled,vlined]{algorithm2e}
\usepackage{algpseudocode}
\usepackage{graphicx}
\usepackage{float} 
\usepackage{subfigure}
\usepackage{subcaption}
\usepackage{stfloats}

\newtheorem{theorem}{Theorem}

\newtheorem{lemma}{Lemma}

\newtheorem{corollary}{Corollary}

\allowdisplaybreaks
\makeatletter
\def\ScaleIfNeeded{%
\ifdim\Gin@nat@width>\linewidth \linewidth \else \Gin@nat@width
\fi } \makeatother

\graphicspath{{Fig/}}

\begin{document}

\title{Movable-Element RIS-Aided Wireless Communications: An Element-Wise Position Optimization Approach}
\author{Jingjing Zhao, Qingyi Huang, Kaiquan Cai, Quan Zhou, Xidong Mu, and Yuanwei Liu
\thanks{J. Zhao, Q. Huang, K. Cai, and Q. Zhou are with the School of Electronics and Information Engineering, Beihang University, Beijing, China, and also with the State Key Laboratory of CNS/ATM, Beijing, China. (e-mail:\{jingjingzhao, qyhuang, caikq, quanzhou\}@buaa.edu.cn). 
X. Mu is with the Centre for Wireless Innovation (CWI), Queen's University Belfast, Belfast, BT3 9DT, U.K. (e-mail: x.mu@qub.ac.uk).
Y. Liu is with the Department of Electrical and Electronic Engineering, the University of Hong Kong, Hong Kong, China (e-mail: yuanwei@hku.hk).}
}

\maketitle
\begin{abstract}
A point-to-point movable element (ME) enabled reconfigurable intelligent surface (ME-RIS) communication system is investigated, where each element position can be flexibly adjusted to create favorable channel conditions. For maximizing the communication rate, an efficient ME position optimization approach is proposed. Specifically, by characterizing the cascaded channel power gain in an element-wise manner, the position of each ME is iteratively updated by invoking the successive convex approximation method. Numerical results unveil that 1) the proposed element-wise ME position optimization algorithm outperforms the gradient descent algorithm; and 2) the ME-RIS significantly improves the communication rate compared to the conventional RIS with fixed-position elements.

\end{abstract}
\section{Introduction}
The reconfigurable intelligent surface (RIS) is a cost-effective technology to build a programmable wireless environment with the customized phase responses of a set of passive elements~\cite{zeng2020reconfigurable}. 
Thanks to the capability of steering signals toward the desired directions, one of RIS applications is to compensate for the propagation path loss and therefore improve the received signal power. Moreover, since no active radio frequency (RF) chain is needed, an RIS consumes much less energy compared to the conventional multiple-antenna technique~\cite{8910627}. 
Note that the conventional RIS adopts fixed-position antennas (FPAs), where the channel variation in the continuous spatial field is not fully leveraged. To further improve the spatial diversity and multiplexing gain, position-adjustable antenna (PAA) techniques, including fluid antenna (FA)~\cite{9264694,9650760} and movable antenna (MA)~\cite{10318061,10354003}, have been proposed recently. Specifically, the position of each antenna can be flexibly adjusted within a confined area, ranging from several to tens of wavelengths, and thereby creating more favorable channel conditions~\cite{10243545}. 

Given the aforementioned benefits, some recent works have started to explore the integration of PAA techniques into RIS-aided communications~\cite{xinwei, 10430366,MA-RIS-YAN}. {In~\cite{xinwei}, the authors investigated the performance gain brought by MAs in the RIS-aided communication system and addressed the sum rate maximization problem, where the MAs positions could be adjusted within a one-dimensional region.} The authors of~\cite{10430366} proposed a unified non-uniform discrete phase shift design for the movable element (ME)-enabled RIS (ME-RIS) communications to eliminate phase offsets and enhance the system performance. In~\cite{MA-RIS-YAN}, the impact of transmit power and number of MEs on the outage probability performance of ME-RIS-aided communications was studied.

The ME position optimization is an essential problem in the ME-RIS-aided communication system, which is intractable to address as the channel vectors/matrices are highly non-convex with respect to (w.r.t.) MEs positions~\cite{10354003}. Currently, the commonly adopted solving approach is the gradient descent algorithm (GDA). However, the main drawback of the GDA is that it is easily trapped into the local optimality given the multimodal problem. To solve this issue, we propose a novel element-wise ME position optimization approach, where position variables for all MEs are decoupled and then iteratively optimized by invoking the successive convex approximation (SCA) method. Our numerical results unveil the superiority of the proposed element-wise ME position algorithm compared to the GDA.   
\vspace{-0.2cm}
\section{System Model and Problem Formulation}
In this section, we present a point-to-point ME-RIS-aided downlink multiple-input single-output (MISO) communication system, and formulate the joint ME position, base station (BS) active beamforming, and ME-RIS passive beamforming  optimization problem. 

\begin{figure}[h]
	\centering
	\includegraphics[scale=0.3]{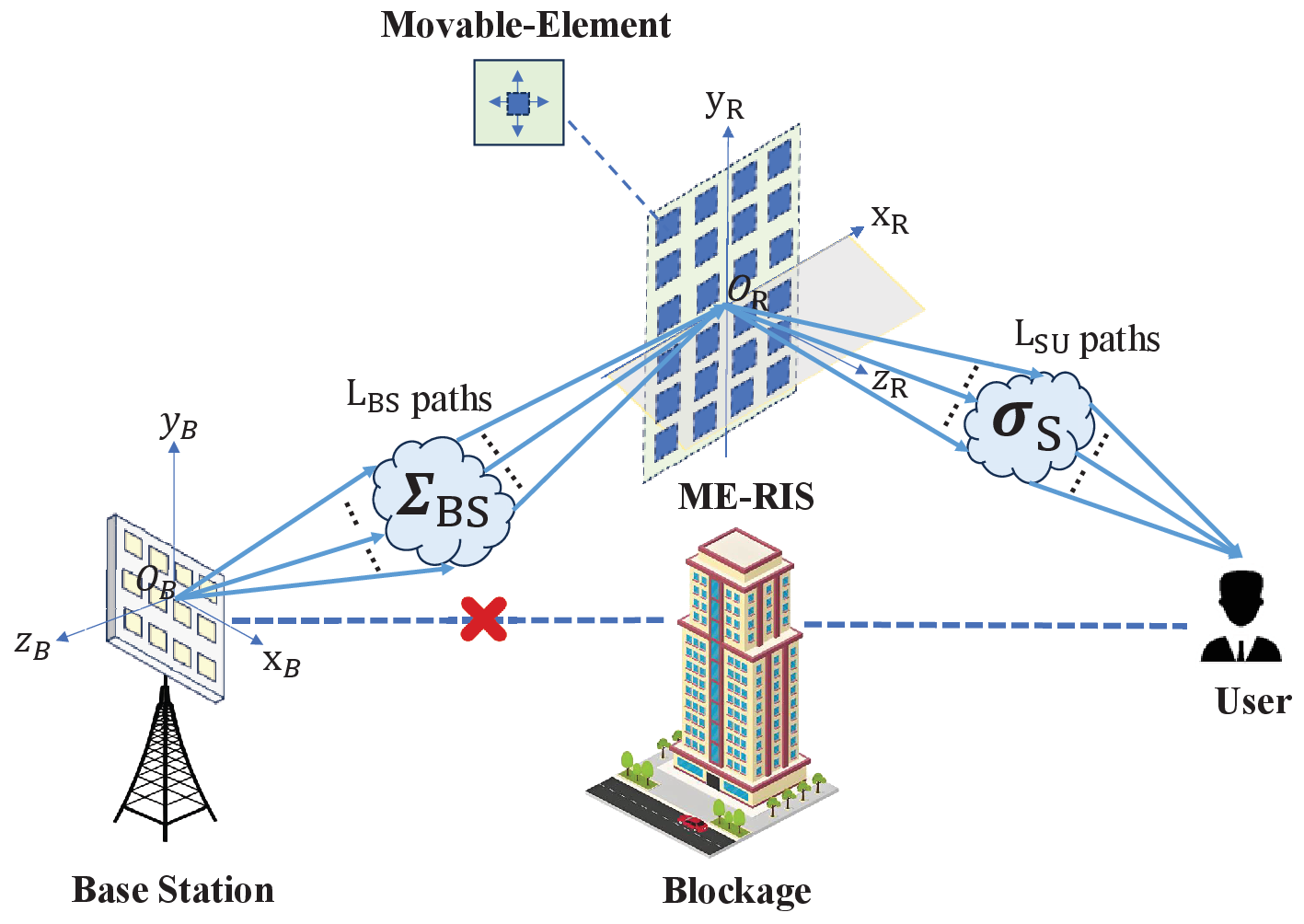}
	\caption{Illustration of the point-to-point ME-RIS-aided downlink MISO communications system.}
	\label{fig:system_model}
\end{figure}
\vspace{-0.4cm}
\subsection{System Model}
As shown in Fig.~\ref{fig:system_model}, we consider a point-to-point ME-RIS-aided MISO communication system, where a BS equipped with $M$ FPAs transmits signals to a single-antenna user. The direct channel from the BS to the user is assumed to be severely blocked, and an ME-RIS with $N$ elements is deployed to provide reflected channel link. We denote the ME-RIS phase-shift matrix as $\mathbf{\Theta}=\text{diag}\left(e^{j\theta_1},e^{j\theta_2}, \dots,  e^{j\theta_N}\right)$, where $\theta_N\in[0,2\pi)$ represents the phase shift applied to the $n$-th ME.
The local coordinate systems $x_{\text{S}}-O_{\text{S}}-y_{\text{S}}$ and $x_{\text{B}}-O_{\text{B}}-y_{\text{B}}$ are considered at the ME-RIS and the BS, respectively, with $O_{\text{S}}=[0,0]^T$ and  $O_{\text{B}}=[0,0]^T$ representing the corresponding reference points. Let $\mathbf{u}_n = \left[x_{n},y_{n}\right]^{T}$ and $\mathbf{r}_m = \left[x_{m},y_{m}\right]^{T}$ denote the position of the $n$-th element at the ME-RIS and that of the $m$-th FPA at the BS, respectively. Therefore, the elements positions matrix (EPM) at the ME-RIS is represented by $\mathbf{U} = \left[\mathbf{u}_1, \mathbf{u}_2, ..., \mathbf{u}_N\right]\in\mathbb{R}^{2\times N}$.

We consider a quasi-static slow-fading channel model, where there exists $L_{\text{BS}}$ and $L_{\text{SU}}$ paths for the BS-RIS and the RIS-user channels, respectively. The field-response vector (FRV) at the $m$-th FPA of the BS is given by
\begin{equation}
    \mathbf{e}\left(\mathbf{r}_m\right) = \left[e^{j\frac{2\pi}{\lambda}\rho_{\text{B}}^{1}\left(\mathbf{r}_m\right)},e^{j\frac{2\pi}{\lambda}\rho_{\text{B}}^{2}\left(\mathbf{r}_m\right)}, ..., e^{j\frac{2\pi}{\lambda}\rho_{\text{B}}^{L_{\text{BS}}}\left(\mathbf{r}_m\right)}\right]^{T},
\end{equation}
where $\rho_{\text{B}}^{j}\left(\mathbf{r}_m\right) = x_m\cos{\theta_{\text{BS,B}}^j}\sin{\phi_{\text{BS,B}}^j}+y_m\sin{\theta_{\text{BS,B}}^j}$ represents the signal propagation difference between the $m$-th FPA and the reference point $O_{\text{B}}$ for the $j$-th path, with $\theta_{\text{BS,B}}^j$ and $\phi_{\text{BS,B}}^j$ denoting the elevation and azimuth angle-of-departure (AoD) of the $j$-th path between the BS and the ME-RIS, respectively. Accordingly, we can express the field response matrix (FRM) for all FPAs as follows:
\begin{equation}
    \mathbf{E}=\left[\mathbf{e}\left(\mathbf{r}_1\right), \mathbf{e}\left(\mathbf{r}_2\right), ..., \mathbf{e}\left(\mathbf{r}_M\right)\right]\in\mathbb{C}^{L_{\text{BS}}\times M},
\end{equation}
which is a constant matrix given the fixed positions of antennas at the BS. 

The FRV of the $n$-th element at the ME-RIS for the incident channel is denoted by
\begin{equation}
    \mathbf{f}_{\text{in}}\left(\mathbf{u}_n\right) = \left[e^{j\frac{2\pi}{\lambda}\rho_{\text{S,in}}^{1}\left(\mathbf{u}_n\right)},e^{j\frac{2\pi}{\lambda}\rho_{\text{S,in}}^{2}\left(\mathbf{u}_n\right)}, ..., e^{j\frac{2\pi}{\lambda}\rho_{\text{S,in}}^{L_{\text{BS}}}\left(\mathbf{u}_n\right)}\right]^{T},
\end{equation}
where $\rho_{\text{S,in}}^j\left(\mathbf{u}_n\right) = x_n\cos{\theta_{\text{BS,S}}^j}\sin{\phi_{\text{BS,S}}^j}+y_n\sin{\theta_{\text{BS,S}}^j}$ denotes the difference of the signal propagation distance between the $n$-th ME and the reference point $O_{\text{S}}$ for the $j$-th path, with $\theta_{\text{BS,S}}^j$ and $\phi_{\text{BS,S}}^j$ representing the elevation and azimuth angle-of-arrival (AoA) of the $j$-th path between the BS and the ME-RIS, respectively. Accordingly, the FRM at the ME-RIS for the incident channel is given by $\mathbf{F}_{\text{in}}\left(\mathbf{U}\right) = \left[\mathbf{f}_{\text{in}}\left(\mathbf{u}_1\right),\mathbf{f}_{\text{in}}\left(\mathbf{u}_2\right),...,\mathbf{f}_{\text{in}}\left(\mathbf{u}_N\right)\right]\in\mathbb{C}^{L_{\text{BS}}\times N}$. Define the path-response matrix (PRM) for all paths from $O_{\text{B}}$ to $O_{\text{S}}$ as $\mathbf{\Sigma}_{\textbf{BS}} \in \mathbb{C}^{L_{\text{BS}}\times L_{\text{BS}}}$, then the BS-RIS channel matrix is given by
$\mathbf{H}\left(\mathbf{U}\right) = \mathbf{F}_{\text{in}}^{H}\left(\mathbf{U}\right)\mathbf{\Sigma}_{\text{BS}}\mathbf{E}$. 
Moreover, the FRV of the $n$-th element at the ME-RIS for the RIS-user channel is given by
\begin{equation}
    \mathbf{f}\left(\mathbf{u}_n\right) = \left[e^{j\frac{2\pi}{\lambda}\rho_{\text{S}}^{1}\left(\mathbf{u}_n\right)},e^{j\frac{2\pi}{\lambda}\rho_{\text{S}}^{2}\left(\mathbf{u}_n\right)}, ..., e^{j\frac{2\pi}{\lambda}\rho_{\text{S}}^{L_{\text{SU}}}\left(\mathbf{u}_n\right)}\right]^{T},
\end{equation}
where $\rho_{\text{S}}^k\left(\mathbf{u}_n\right) = x_n\cos{\theta_{\text{SU}}^k}\sin{\phi_{\text{SU}}^k}+y_n\sin{\theta_{\text{SU}}^k}$, with $\theta_{\text{SU}}^k$ and  $\phi_{\text{SU}}^k$ representing the elevation and azimuth AoD for the $k$-th path between the ME-RIS and the user, respectively. With $\mathbf{F}\left(\mathbf{U}\right) = \left[\mathbf{f}\left(\mathbf{u}_1\right),\mathbf{f}\left(\mathbf{u}_2\right),...,\mathbf{f}\left(\mathbf{u}_N\right)\right]\in\mathbb{C}^{L_{\text{SU}}\times N}$, we have the RIS-user channel response as $\mathbf{g}\left(\mathbf{U}\right) = \mathbf{\sigma}_{\text{S}}\mathbf{F}\left(\mathbf{U}\right)$, where $\mathbf{\sigma}_{\text{S}}\in\mathbb{C}^{1\times L_{\text{SU}}}$ denotes the path response from $O_{\text{S}}$ to the user.

The signal received by the user is given by
\begin{align}
    y = & \mathbf{g}\left({\mathbf{U}}\right)\mathbf{\Theta}\mathbf{H}\left({\mathbf{U}}\right)\mathbf{w}x + n,
\end{align} 
{where $x$ is the transmitted signal with $\mathbb{E}\left[\left|x\right|^2\right]=1$, $\mathbf{w}\in\mathbb{C}^{M \times 1}$ represents the beamforming vector at the BS, and $n \sim\mathcal{CN}\left(0, \sigma^{2}\right)$ denotes the additive white Gaussian noise at the user.} The user’s communication rate is given by
\begin{equation}
\label{eq:data-rate}
R = \log_2\left(1 + \frac{\left|\mathbf{g}\left({\mathbf{U}}\right)\mathbf{\Theta}\mathbf{H}\left({\mathbf{U}}\right)\mathbf{w}\right|^2}{ \sigma^2}\right).
\end{equation}

\subsection{Problem Formulation}
In this work, we aim at maximizing the point-to-point communication rate by jointly optimizing the MEs positions $\mathbf{U}$, the BS active beamforming $\mathbf{w}$, and the ME-RIS passive beamforming $\mathbf{\Theta}$, while satisfying MEs positions constraints and feasible constraints of reflection coefficients. The resulting optimization problem is formulated as
\begin{subequations}
\label{eq:optimization_problem1}
\begin{equation}
\label{eq:objective-function1}
    \max_{\mathbf{U},\mathbf{w}, \mathbf{\Theta}} \left|\mathbf{g}\left({\mathbf{U}}\right)\mathbf{\Theta}\mathbf{H}\left({\mathbf{U}}\right)\mathbf{w}\right|^2,
\end{equation}
\begin{equation}
\label{eq:move-region}
    {\rm{s.t.}} \ \ \mathbf{u}_n\in\mathcal{C}, 1\leq n\leq N,
\end{equation}
\begin{equation}
\label{eq:ME-minimum-distance}
    \left\|\mathbf{u}_n - \mathbf{u}_{n'}\right\|_2\geq D_0, 1\leq n\neq n'\leq N,
\end{equation}
\begin{equation}
    \label{eq:BS-transmit-power-constraint1}
  {\rm{Tr}}\left(\mathbf{w}^{H}\mathbf{w}\right)\leq P_{\text{max}}, 
\end{equation}
\begin{equation}
    \label{eq:passive-beamforming-phase-constraint1}
    \theta_{n} \in [0,2\pi), \forall 1\leq n\leq N, 
\end{equation}
\end{subequations}
where \eqref{eq:move-region} confines that MEs should be located in the feasible moving region $\mathcal{C}$, \eqref{eq:ME-minimum-distance} restricts the minimum distance $D_0$ among MEs for avoiding the coupling effect, \eqref{eq:BS-transmit-power-constraint1} ensures that the BS maximum transmit power does not exceed $P_{\text{max}}$, and \eqref{eq:passive-beamforming-phase-constraint1} gives the phase adjustment constraints at the ME-RIS. Problem~\eqref{eq:optimization_problem1} is intractable to solve due to the tight coupling among $\mathbf{U}$, $\mathbf{w}$, and $\mathbf{\Theta}$ as well as the high non-convexity w.r.t. the involved variables. To solve these challenges, we propose an alternative optimization (AO)-based method for decoupling the original problem in the following. 

\section{Proposed Solution}

For optimizing the BS active beamforming $\mathbf{w}$, the optimal solution can be obtained with the maximum-ratio transmission strategy (MRT)~\cite{8811733}, i.e.,
\begin{equation}
    \mathbf{w^*}=\sqrt{P_{\text{max}}}\frac{\left(\mathbf{g}\left(\mathbf{U}\right)\mathbf{\Theta H}\left(\mathbf{U}\right)\right)^{H}}{\left\|\mathbf{g}\left(\mathbf{U}\right)\mathbf{\Theta H}\left(\mathbf{U}\right)\right\|}.
\end{equation}
By substituting $\mathbf{w^*}$ to \eqref{eq:objective-function1}, the problem~\eqref{eq:optimization_problem1} can be transformed into the following problem
\begin{subequations}
\label{eq:optimization_problem2}
\begin{equation}
\label{eq:objective-function2}
    \max_{\mathbf{U}, \mathbf{\Theta}} \left\|\mathbf{g}\left({\mathbf{U}}\right)\mathbf{\Theta}\mathbf{H}\left({\mathbf{U}}\right)\right\|^2,
\end{equation}
\begin{equation}
    {\rm{s.t.}} \ \ \eqref{eq:move-region}, \eqref{eq:ME-minimum-distance}, \eqref{eq:passive-beamforming-phase-constraint1}.
\end{equation}
\end{subequations}

\begin{figure}[h]
	\centering
	\includegraphics[scale=0.35]{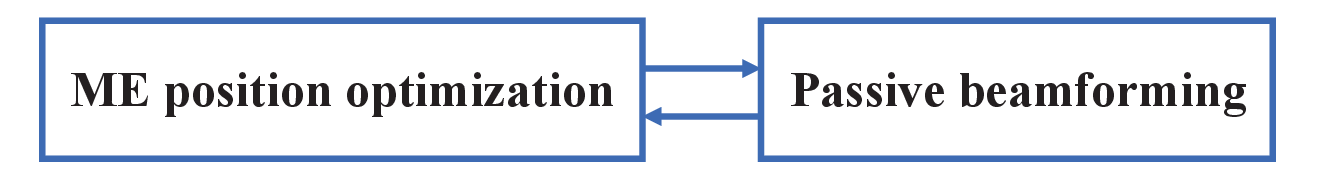}
	\caption{Illustration of the AO method.}
	\label{fig:AO} 
\end{figure}

By invoking the AO method, the problem~\eqref{eq:optimization_problem2} can be decomposed into two subproblems, i.e., the ME position optimization and the ME-RIS passive beamforming, as shown in Fig.~\ref{fig:AO}. Existing works have demonstrated that SCA~\cite{10243545} or manifold optimization~\cite{10542230} methods can effectively solve the RIS passive beamforming subproblem, for which the details are omitted here for the ease of brevity. In the following, we focus on solving the ME position optimization subproblem.

To start with, we first transform problem~\eqref{eq:optimization_problem2} into a more tractable form. To facilitate the design, we define the phase-shift vectors of the ME-RIS as $\mathbf{q}=\left[e^{j\theta_1},e^{j\theta_2}, ..., e^{j\theta_N}\right]^{H}$. Then we have $\left\|\mathbf{g}\left(\mathbf{U}\right)\mathbf{\Theta}\mathbf{H}\left(\mathbf{U}\right)\right\|^2 = \left\|\mathbf{q}^H\mathbf{V}\left(\mathbf{U}\right)\right\|^2$, where $\mathbf{V}\left(\mathbf{U}\right)\in\mathbb{C}^{N\times M}$ is given by
\begin{align}
\label{eq:cas-channel}
    \mathbf{V}\left(\mathbf{U}\right) & = \text{diag}\left(\mathbf{g}\left(\mathbf{U}\right)\right)\mathbf{H}\left(\mathbf{U}\right)\nonumber \\
    & = \text{diag}\left(\mathbf{\sigma}_{\text{S}}\mathbf{F}\left(\mathbf{U}\right)\right)\mathbf{F}_{\text{in}}^{{H}}\left(\mathbf{U}\right)\mathbf{\Sigma}_{\text{BS}}\mathbf{E}
   \nonumber\\ 
   & =\left(\mathbf{\sigma}_{\text{S}}\mathbf{F}\left(\mathbf{u}_1\right)\mathbf{\Sigma}_{\text{BS}}\mathbf{E}, ..., \mathbf{\sigma}_{\text{S}}\mathbf{F}\left(\mathbf{u}_N\right)\mathbf{\Sigma}_{\text{BS}}\mathbf{E}\right)^T.
\end{align}
In~\eqref{eq:cas-channel}, $\mathbf{F}\left(\mathbf{u}_n\right)=\mathbf{f}\left(\mathbf{u}_n\right)\left(\mathbf{f}_{\text{in}}\left(\mathbf{u}_n\right)\right)^H\in\mathbb{C}^{L_{\text{SU}}\times L_{\text{BS}}}$ is only determined by the position of the $n$-th ME.
Accordingly,  $\left\|\mathbf{q}^H\mathbf{V}\left(\mathbf{U}\right)\right\|^2$ can be rewritten in an element-wise manner as follows:
\begin{align} 
\label{eq:element-wise}\left\|\mathbf{q}^H\mathbf{V}\left(\mathbf{U}\right)\right\|^2 = \left\|{\mathbf{a}_{n}\mathbf{F}\left(\mathbf{u}_n\right)\mathbf{b} + {\sum_{n'\neq n}^N \mathbf{a}_{n'}\mathbf{F}\left(\mathbf{u}_{n'}\right)\mathbf{b}}}\right\|^2,
\end{align}
where $\mathbf{a}_{n} = \mathbf{q}^H[n]\mathbf{\sigma}_{\text{S}} \in\mathbb{C}^{1\times L_{\text{SU}}}$, $\mathbf{a}_{n'} = \mathbf{q}^H[n']\mathbf{\sigma}_{\text{S}} \in\mathbb{C}^{1\times L_{\text{SU}}}$, and $\mathbf{b} = \mathbf{\Sigma}_{\text{BS}}\mathbf{E}\in\mathbb{C}^{L_{\text{BS}}\times M}$ are constant variables irrelevant to $\mathbf{U}$. Note that the maximization of the achievable rate should jointly consider the optimization of the $N$ ME positions, which is an intractable task due to the tight coupling of $\left\{\mathbf{u}_n\right\}_{n=1}^N$. Thanks to the element-wise derivations in~\eqref{eq:element-wise}, we are inspired to solve $\mathbf{U}$ in an iterative manner, where each $\mathbf{u}_n, n\in\left\{1, ..., N\right\}$ is optimized alternatively with $\mathbf{u}_{n'}, n'\neq n$ being fixed. 

The subproblem for solving $\mathbf{u}_n$ can be given as
\begin{subequations}
\label{eq:element-wise-subproblem}
\begin{equation}
\label{eq:R-objective-function}
	\max_{\mathbf{u}_n} \left\|\mathbf{q}^H\mathbf{V}\left(\mathbf{u}_n\right)\right\|^2,
\end{equation}
\begin{equation}
    \label{eq:C4}
    {\rm{s.t.}} \ \ \ \mathbf{u}_n \in \mathcal{C}, 1\leq n\leq N,
\end{equation}
\begin{equation}
    \label{eq:C5}
    \left\|\mathbf{u}_n - \mathbf{u}_{n'}\right\|_2 \geq D_0, 1\leq n'\leq N, n'\neq n.
\end{equation}
\end{subequations}
Problem~\eqref{eq:element-wise-subproblem} is still intractable to solve because of the non-convex constraints~\eqref{eq:R-objective-function} and~\eqref{eq:C5}. To address this issue, we propose the SCA-based algorithm.
Specifically, for the non-convex objective function~\eqref{eq:R-objective-function}, it is difficult to construct a surrogate function using only the first-order Taylor expansion. Therefore, we propose to construct a quadratic surrogate function by using the second-order Taylor expansion. Specifically, denote $\upsilon\left(\mathbf{u}_n\right) = \left\|\mathbf{q}^H\mathbf{V}\left(\mathbf{u}_n\right)\right\|^2$. The gradient vector and Hessian matrix of $\upsilon\left(\mathbf{u}_n\right)$ over $\mathbf{u}_n$ are denoted by $\nabla \upsilon\left(\mathbf{u}_n\right)\in\mathbb{R}^2$ and $\nabla^2\upsilon\left(\mathbf{u}_n\right)\in\mathbb{R}^{2\times 2}$, respectively, for which the detailed derivations can be found in Appendix~\ref{app:A}. Then, by constructing a positive real number $\xi_{n}$ that satisfies $\xi_{n} \mathbf{I}_2 \succeq \nabla^2\upsilon\left(\mathbf{u}_n\right)$, the quadratic surrogate function that globally lower bounds $\upsilon\left(\mathbf{u}_n\right)$ with given points $\left(\mathbf{u}_n^{\left(i\right)}, {\upsilon}\left(\mathbf{u}_n^{\left(i\right)}\right)\right)$ can be constructed as
\begin{align}
\label{eq:mu-lower-bound}
    \bar{\upsilon}\left(\mathbf{u}_n\right) = & \  {\upsilon}\left(\mathbf{u}_n^{\left(i\right)}\right) + \nabla{\upsilon}\left(\mathbf{u}_n^{\left(i\right)}\right)\left(\mathbf{u}_n - \mathbf{u}_n^{\left(i\right)}\right)\nonumber\\
    &- \frac{\xi_{n}}{2}\left(\mathbf{u}_n - \mathbf{u}_n^{\left(i\right)}\right)^T\left(\mathbf{u}_n - \mathbf{u}_n^{\left(i\right)}\right).
\end{align}
For the Hessian matrix $\nabla^2\upsilon\left(\mathbf{u}_n\right)$, we have $\left\|\nabla^2\upsilon\left(\mathbf{u}_n\right)\right\|_2\mathbf{I}_2\succeq\nabla^2\upsilon\left(\mathbf{u}_n\right)$ and $
\left\|\nabla^2\upsilon\left(\mathbf{u}_n\right)\right\|_{\text{F}} \geq \left\|\nabla^2\upsilon\left(\mathbf{u}_n\right)\right\|_2$. 
As such, $\left\|\nabla^2\upsilon\left(\mathbf{u}_n\right)\right\|_{\text{F}}\mathbf{I}_2\succeq\nabla^2\upsilon\left(\mathbf{u}_n\right)$ is guaranteed. We can thus set $\xi_{n}$ as in \eqref{eq:xi-constant}, which satisfies $\xi_{n} \geq \left\|\nabla^2\upsilon\left(\mathbf{u}_n\right)\right\|_{\text{F}} \geq \left\|\nabla^2\upsilon\left(\mathbf{u}_n\right)\right\|_2$, and therefore $\xi_{n} \mathbf{I}_2 \succeq \nabla^2\upsilon\left(\mathbf{u}_n\right)$. 

\begin{figure*}[!h]
\begin{align}
\label{eq:xi-constant}
    \xi_{n} = \frac{8\pi^2\left|c_{n}\right|}{\lambda^2}\sqrt{\left( \sum_{j=1}^{L_{\text{BS}}}\sum_{k=1}^{L_{\text{SU}}}\left(\zeta^{k,j}\right)^2|a_{n}^k| |b^j|\right)^2 + 2\left( \sum_{j=1}^{L_{\text{BS}}}\sum_{k=1}^{L_{\text{SU}}}\zeta^{k,j}\mu^{k,j}|a_{n}^k| |b^j|\right)^2 + \left( \sum_{j=1}^{L_{\text{BS}}}\sum_{k=1}^{L_{\text{SU}}}\left(\mu^{k,j}\right)^2|a_{n}^k| |b^j|\right)^2},
\end{align}
\hrulefill
\end{figure*}

For the non-convex constraint~\eqref{eq:C5}, we can construct the globally lower bound of $\| \mathbf{u}_n -\mathbf{u}_{n'} \|_2$ with the first-order Taylor expansion, given that $\| \mathbf{u}_n -\mathbf{u}_{n'} \|_2$ is a convex function w.r.t. $\mathbf{u}_n$. As such, in the $i$-th iteration of the SCA, \eqref{eq:C5} can be equivalently transformed into
\begin{align}
\label{eq:distance-relax}
&\frac{1}{\| \mathbf{u}_n^{\left(i\right)} - \mathbf{u}_{n'}\|_2} \left(\mathbf{u}_n^{\left(i\right)} - \mathbf{u}_{n'}\right)^T \left(\mathbf{u}_n - \mathbf{u}_{n'}\right) \geq D_0, \nonumber\\
& \ \ \ \ 1\leq n'\leq N, n'\neq n.
\end{align}
As a result, in the $i$-th iteration of the SCA, problem~\eqref{eq:element-wise-subproblem} can be relaxed as
\begin{subequations}
\label{eq:convex-problem}
\begin{equation}
	\max_{\mathbf{u}_n} \bar{\upsilon}\left(\mathbf{u}_n\right),
\end{equation}
\begin{equation}
   {\rm{s.t.}} \ \ \eqref{eq:C4}, \eqref{eq:distance-relax}.
\end{equation}
\end{subequations}
Problem~\eqref{eq:convex-problem} is a quadratic programming (QP) problem and can be directly solved with standard convex problem solvers such as CVX~\cite{cvx}. 
\begin{figure*}[!h]
	\begin{minipage}{0.32\linewidth}		\centerline{\includegraphics[width=\textwidth]{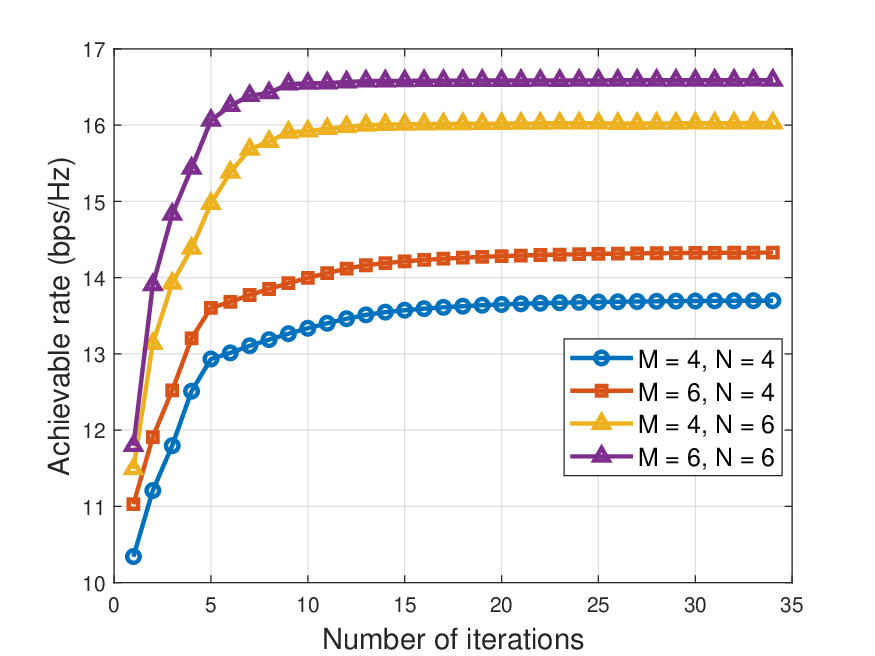}}
		\caption{Convergence performance of the AO algorithm, with $A=5\lambda$.}
        \label{fig:convergence_performance}
	\end{minipage}
	\begin{minipage}{0.32\linewidth}
		\centerline{\includegraphics[width=\textwidth]{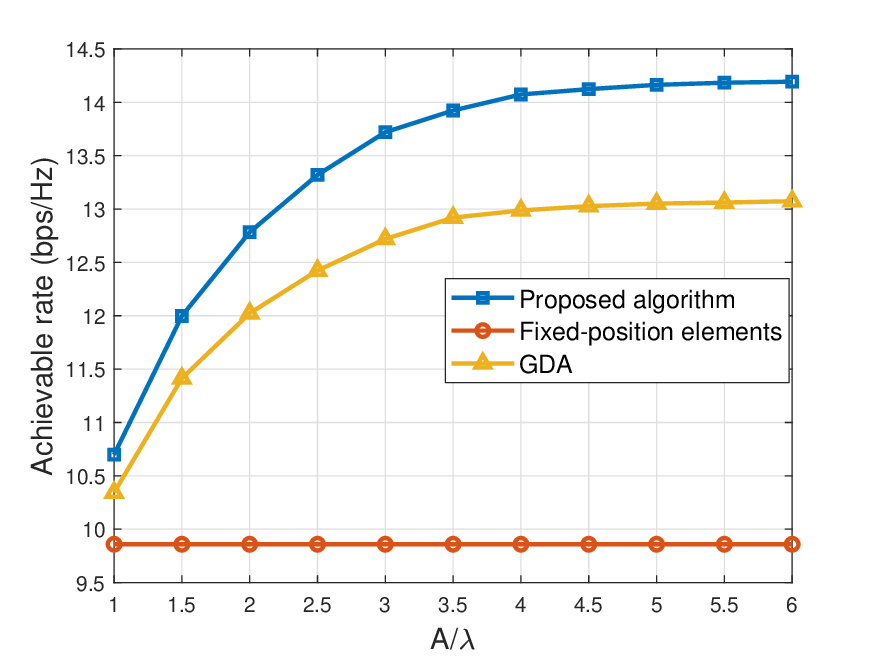}}
		\caption{Communication rate versus MEs moving region size, with $M=N=4$.}
        \label{fig:rate_region_size}
	\end{minipage}
	\begin{minipage}{0.32\linewidth}
		\centerline{\includegraphics[width=\textwidth]{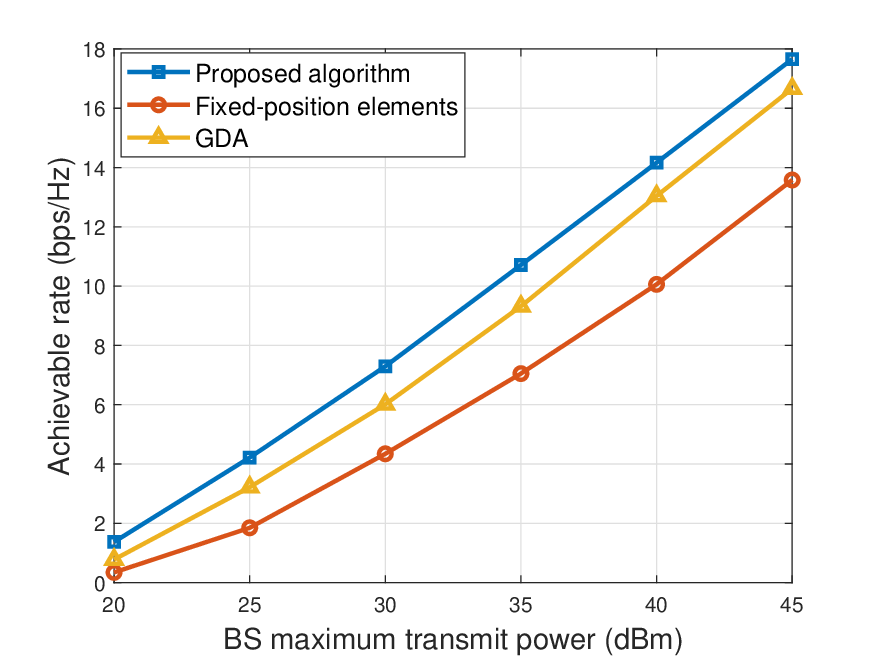}}
		\caption{Communication rate versus BS maximum transmit power, with $M=N=4$.}
        \label{fig:rate_SNR}
	\end{minipage}
\end{figure*}
\begin{algorithm}[h]
\small
\caption{ME Position Optimization Algorithm}
\label{alg:U-optimization}
\LinesNumbered
\KwIn{$\{\mathbf{r}_m\}_{m=1}^M$,$\mathbf{\Sigma}_{\text{BS}}$, $\left\{\mathbf{\sigma}_{\text{S}}\right\}$, $\left\{\theta_{\text{BS,B}}^j\right\}_{j=1}^{L_{\text{BS}}}$, $\{\phi_{\text{BS,B}}^j\}_{j=1}^{L_{\text{BS}}}$, $\{\theta_{\text{BS,S}}^j\}_{j=1}^{L_{\text{BS}}}$, $\{\phi_{\text{BS,S}}^j\}_{j=1}^{L_{\text{BS}}}$, $\{\theta_{\text{SU}}^k\}_{k=1}^{L_{\text{SU}}}$, $\{\phi_{\text{SU}}^k\}_{k=1}^{L_{\text{SU}}}$, $\mathcal{C}$, ${D}_0$, $\mathbf{w}$, $\mathbf{\Theta}$}
\KwOut{$\mathbf{U}$}
Initialize $\left\{\mathbf{u}_n\right\}_{n=1}^N = \left\{\mathbf{u}_n^{\left(0\right)}\right\}_{n=1}^N$;\\
\For{$n = 1\rightarrow{N}$}{
    Set iteration index $i = 0$;\\
    \While{Increment of the achievable rate is above a threshold $\epsilon>0$}{
        Solve problem~\eqref{eq:convex-problem} to obtain $\mathbf{u}_n^{\left(i+1\right)}$;\\
        Update $\mathbf{u}_n = \mathbf{u}_n^{\left(i+1\right)}$;\\
        $i = i+1$;
    }  
}
\end{algorithm}

The overall element-wise optimization algorithm for solving $\mathbf{U}$ is given in \textbf{Algorithm~\ref{alg:U-optimization}}, which consists of an outer loop and an inner loop. Specifically, the $N$ MEs positions are optimized alternatively in the outer loop, while the problem~\eqref{eq:convex-problem} is iteratively solved in the inner loop until the increment of achievable rate is below a threshold $\epsilon$. Since the achievable rate is non-decreasing by solving the problem~\eqref{eq:convex-problem} in each iteration of the inner loop, the convergence of the proposed algorithm is guaranteed given that the communication capacity is upper bounded. The computational complexity of \textbf{Algorithm~\ref{alg:U-optimization}} is analyzed as follows. {For the calculation of $\mathbf{a}_{n}$, $\mathbf{b}$, $c_{n}$, $\nabla \upsilon\left(\mathbf{u}_n\right)$, and $\xi_{n}$, the computational complexity is given by $\mathcal{O}\left(L_{\text{SU}}\right)$, $\mathcal{O}\left(ML_{\text{BS}}\right)$, $\mathcal{O}\left(NL_{\text{BS}}L_{\text{SU}}\right)$, $\mathcal{O}\left(L_{\text{BS}}L_{\text{SU}}\right)$, and $\mathcal{O}\left(L_{\text{BS}}L_{\text{SU}}\right)$, respectively.} Therefore, the complexity for solving problem~\eqref{eq:convex-problem} is $\mathcal{O}\left(N^{1.5}\right)$ if the interior point method is employed~\cite{{boyd2004convex}}.  Therefore, the overall computational complexity of \textbf{Algorithm~\ref{alg:U-optimization}} is $\mathcal{O}\left(NI_{\text{in}}\left(N^{1.5} + NL_{\text{BS}}L_{\text{SU}}\right)\right)$, where $I_{\text{in}}$ is the number of inner iterations for the convergence of the SCA. As can be observed, the computational complexity is polynomial in $N$, $L_{\text{BS}}$, and $L_{\text{SU}}$.

\section{Simulation Results}
In this section, we validate the effectiveness of the proposed AO algorithm through numerical simulations. Referring to the $x_{\text{S}}-y_{\text{S}}-z_{\text{S}}$ coordinate in Fig.~\ref{fig:system_model}, the BS is set at $\left(-10,-5,10\right)$~meters, while the user is randomly distributed within a square region centered at $\left(20,-10,20\right)$~meters with the edge size of $40$~meters. The MEs moving region $\mathcal{C}$ is defined as a square area with size $A\times A$. We assume that the number of paths is the same for all the considered channels, i.e, $L_{\text{BS}} = L_{\text{SU}}=L$. Moreover, we set $P_{\text{max}} = 40$ dBm, while the noise power at the user is set to $\sigma^{2} = -80$ dBm. The involved path angles, i.e.,  $\theta_{\text{BS,B}}^j$, $\phi_{\text{BS,B}}^j$, $\theta_{\text{BS,S}}^j$, $\phi_{\text{BS,S}}^j$, $\theta_{\text{SU}}^k$,  $\phi_{\text{SU}}^k, \forall j,k$, are generated randomly within the range $\left[-\pi/2,\pi/2\right]$. The PRM $\mathbf{\Sigma}_{\textbf{BS}}$ is assumed to be diagonal with each diagonal element following the circularly symmetric
complex Gaussian (CSCG) distribution $\mathcal{CN}\left(0, \beta_\text{BS}/L\right)$, where $\beta_{\text{BS}}=\beta_{0}d_{\text{BS}}^{-\alpha_{0}}$ is the BS-RIS channel power gain, $\beta_{0} = -30$ dB is the channel power gain at the reference distance of $1~\mathrm{m}$, $d_{\text{BS}}$ represents the distance between the BS and the ME-RIS, and $\alpha_{0} = 2.2$ is the path-loss exponent. Moreover, each element of $\mathbf{\sigma}_{\text{S}}$ follows the CSCG distribution $\mathcal{CN}\left(0, \beta_\text{S}/L\right)$, where $\beta_\text{S}=\beta_{0}d_{\text{S}}^{-\alpha_{0}}$ with $d_{\text{S}}$ representing the distance between the RIS and the user. For the simulations, the penalty-based SCA method proposed in~\cite{9570143} is utilized for the ME-RIS beamforming design embedded within the AO algorithm.
The simulation results are averaged over $1000$ user distributions and channel realizations.

In Fig.~\ref{fig:convergence_performance}, we study the convergence performance of the AO algorithm, with $A=5\lambda$.  One can first observe that the communication rate keeps increasing and converges to steady points after around $25$ iterations for all the given $M$ and $N$ values. Moreover, it is also seen that the communication rate increases with larger $M$ and $N$. This is expected, since larger $M$ and $N$ can bring in higher power gain. 

In Fig.~\ref{fig:rate_region_size}, we show the communication rate versus the MEs moving region size $A$, where we set $M=N=4$. It can be observed that the proposed element-wise position optimization algorithm outperforms the GDA, which is susceptible to local optima. Moreover, the ME-RIS shows its superiority compared to the counterpart with fixed-position elements (FPEs). The performance gap increases with larger $A$ and gets saturated after around $A/\lambda=5$. This is because, the communication rate reaches to the upper bound when there exists sufficient optimal positions for a given $N$, which means that enlarging $A$ can not bring in more performance gain. We further depict the achievable rate versus the BS maximum transmit power $P_{\text{max}}$ in Fig.~\ref{fig:rate_SNR}, where we set $A=5\lambda$. The proposed algorithm is shown to outperform the GDA and the FPEs benchmarks under all considered $P_{\text{max}}$ values.

\section{Conclusions}
A novel element-wise ME position optimization method has been proposed for the ME-RIS-aided communication system, where position variables for all MEs are decoupled and iteratively optimized for maximizing the communication rate. It showed that the ME-RIS significantly outperforms the RIS with FPEs. Moreover, the proposed algorithm shows its superiority over the conventional GDA for ME position optimization.

\begin{appendices} 
\section{Derivation of $\nabla \upsilon\left(\mathbf{u}_n\right)$ and $\nabla^2\upsilon\left(\mathbf{u}_n\right)$}
\label{app:A}
For ease of brevity, we define constant $c_{n}$ as
\begin{equation}
    c_{n} = \left|c_{n}\right|e^{j\angle c_{n}} \triangleq \sum_{n'\neq n}^N \mathbf{a}_{n'}\mathbf{F}\left(\mathbf{u}_{n'}\right)\mathbf{b},
\end{equation}
where $|c_{n}|$ and $\angle c_{n}$ represent the amplitude and phase of $c_{n}$, respectively.
Further denote the $k$-th element of $\mathbf{a}_{n}$ and the $j$-th element of $\mathbf{b}$ as ${a}_{n}^k = \left|a_{n}^k\right|e^{j\angle a_{n}^k}$ and ${b}^j = \left|{b}^j\right|e^{j\angle {b}^j}$, respectively, with the $a_{n}^k$ amplitude $\left|a_{n}^k\right|$ and phase $\angle a_{n}^k$, as well as the ${b}^j$ amplitude $\left|{b}^j\right|$ and phase $\angle {b}^j$. According to \eqref{eq:element-wise}, $\mathbf\upsilon\left(\mathbf{u}_n\right)$ can be expressed as in~\eqref{eq:y-r-n}, where $\phi^{k,j}\left(\mathbf{u}_n\right)$ is defined as in~\eqref{eq:phi}. 
\begin{figure*}[hb]
    \hrulefill
    \begin{align}
    \label{eq:y-r-n}
        \upsilon\left(\mathbf{u}_n\right) & \ = \left|\mathbf{a}_{n}\mathbf{F}\left(\mathbf{u}_n\right)\mathbf{b} + c_{n}\right|^2\nonumber\\
     & \ = \left|\sum_{j=1}^{L_{\text{BS}}}\sum_{k=1}^{L_{\text{SU}}}\left|a_{n}^k\right|\left|{b}^j\right|e^{j\left[\frac{2\pi}{\lambda}\left(\rho_{\text{S}}^k\left(\mathbf{u}_n\right)-\rho_{\text{S,in}}^j\left(\mathbf{u}_n\right)\right)+\angle b^j +\angle a_{n}^k\right]}+|c_{n}|e^{j\angle c_{n}}\right|^2\nonumber \\
     & \ = \left(\sum_{j=1}^{L_{\text{BS}}}\sum_{k=1}^{L_{\text{SU}}}\left|a_{n}^k\right|\left|{b}^j\right|\cos \phi^{k,j}\left(\mathbf{u}_n\right)+\left|c_{n}\right|\cos \angle c_{n}\right)^2 + \left(\sum_{j=1}^{L_{\text{BS}}}\sum_{k=1}^{L_{\text{SU}}}\left|a_{n}^k\right|\left|{b}^j\right|\sin \phi^{k,j}\left(\mathbf{u}_n\right)+\left|c_{n}\right|\sin \angle c_{n}\right)^2,
\end{align}
\end{figure*}
\begin{figure*}[hb]
    \begin{align}
    \label{eq:phi}
    \phi^{k,j}\left(\mathbf{u}_n\right) = \frac{2\pi}{\lambda}\left[\left(\underbrace{\cos{\theta_{\text{SU}}^k}\sin{\phi_{\text{SU}}^k} - \cos{\theta_{\text{BS,S}}^j}\sin{\phi_{\text{BS,S}}^j}}_{\zeta^{k,j}}\right)x_n + \left(\underbrace{\sin{\theta_{\text{SU}}^k}-\sin{\theta_{\text{BS,S}}^j}}_{\mu^{k,j}}\right)y_n\right]+\angle b^j +\angle a_{n}^k.
\end{align}
\end{figure*}
The gradient vector of $\upsilon\left(\mathbf{u}_n\right)$ w.r.t. $\mathbf{u}_n$ is given by 
\begin{equation}
    \nabla \upsilon\left(\mathbf{u}_n\right) = \left[\begin{matrix}
        \frac{\partial \upsilon\left(\mathbf{u}_n\right)}{\partial x_n} & \frac{\partial \upsilon\left(\mathbf{u}_n\right)}{\partial y_n}
    \end{matrix}\right],  
\end{equation}
where $\frac{\partial \upsilon\left(\mathbf{u}_n\right)}{\partial x_n}$ and $\frac{\partial \upsilon\left(\mathbf{u}_n\right)}{\partial y_n}$ are given in~\eqref{eq:gradient-vector}.
    \begin{subequations}
    \label{eq:gradient-vector}
        \begin{align}
             \frac{\partial \upsilon\left(\mathbf{u}_n\right)}{\partial x_n} = \frac{4\pi\left|c_{n}\right|}{\lambda}\sum_{j=1}^{L_{\text{BS}}}\sum_{k=1}^{L_{\text{SU}}}\zeta^{k,j}|a_{n}^k||b^j|\sin \left(\angle c_{n} - \phi^{k,j}\left(\mathbf{u}_n\right)\right),
        \end{align}
        \begin{align}
            \frac{\partial \upsilon\left(\mathbf{u}_n\right)}{\partial y_n} = \frac{4\pi\left|c_{n}\right|}{\lambda} \sum_{j=1}^{L_{\text{BS}}}\sum_{k=1}^{L_{\text{SU}}}\mu^{k,j}|a_{n}^k||b^j|\sin \left(\angle c_{n} - \phi^{k,j}\left(\mathbf{u}_n\right)\right).
        \end{align}
    \end{subequations}

Moreover, the Hessian matrix of $\upsilon\left(\mathbf{u}_n\right)$ w.r.t. $\mathbf{u}_n$ is given by
\begin{equation}
\label{eq:mu-Hessian}
    \nabla^2 \upsilon\left(\mathbf{u}_n\right) = \left[\begin{matrix}
        \frac{\partial^2 \upsilon\left(\mathbf{u}_n\right)}{\partial x_n \partial x_n} & \frac{\partial^2 \upsilon\left(\mathbf{u}_n\right)}{\partial x_n \partial y_n} \\
        \frac{\partial^2 \upsilon\left(\mathbf{u}_n\right)}{\partial y_n \partial x_n} & \frac{\partial^2 \upsilon\left(\mathbf{u}_n\right)}{\partial y_n \partial y_n}
    \end{matrix}
    \right],  
\end{equation}
where $\frac{\partial^2 \upsilon\left(\mathbf{u}_n\right)}{\partial x_n \partial x_n}$, $\frac{\partial^2 \upsilon\left(\mathbf{u}_n\right)}{\partial x_n \partial y_n}$, $\frac{\partial^2 \upsilon\left(\mathbf{u}_n\right)}{\partial y_n \partial x_n}$, and $\frac{\partial^2 \upsilon\left(\mathbf{u}_n\right)}{\partial y_n \partial y_n}$ are given in~\eqref{eq:Hessian-elements}.

\begin{subequations}
		\label{eq:Hessian-elements}
		\begin{equation}
		\begin{aligned}
				&\frac{\partial^2 \upsilon\left(\mathbf{u}_n\right)}{\partial x_n^2} \\
                &= -\frac{8\pi^2\left|c_{n}\right|}{\lambda^2} \sum_{j=1}^{L_{\text{BS}}}\sum_{k=1}^{L_{\text{SU}}} \left(\zeta^{k,j}\right)^2 |a_{n}^k| |b^j|\cos \left(\angle c_{n}-\phi^{k,j}\left(\mathbf{u}_n\right) \right),
		\end{aligned}
		\end{equation}
        \begin{equation}
            \begin{aligned}
            &\frac{\partial^2 \upsilon\left(\mathbf{u}_n\right)}{\partial x_n \partial y_n} 
            = \frac{\partial^2 \upsilon\left(\mathbf{u}_n\right)}{\partial y_n \partial x_n} \\
            &= \frac{8\pi^2 \left|c_{n}\right|}{\lambda^2} 
            \sum_{j=1}^{L_{\text{BS}}} \sum_{k=1}^{L_{\text{SU}}} 
            \zeta^{k,j} \mu^{k,j} |a_{n}^{k}| |b^{j}| 
            \cos \left(\angle c_{n} - \phi^{k,j}\left(\mathbf{u}_n\right) \right),
            \end{aligned}
        \end{equation}
		\begin{equation}
			\begin{aligned}
			&\frac{\partial^2 \upsilon\left(\mathbf{u}_n\right)}{\partial y_n^2} \\
            &= -\frac{8\pi^2\left|c_{n}\right|}{\lambda^2} \sum_{j=1}^{L_{\text{BS}}}\sum_{k=1}^{L_{\text{SU}}} \left(\mu^{k,j}\right)^2|a_{n}^k| |b^j|\cos \left(\angle c_{n}-\phi^{k,j}\left(\mathbf{u}_n\right) \right).
			\end{aligned}
		\end{equation}
\end{subequations}
\vspace{-0.7cm}

\end{appendices}

\bibliographystyle{IEEEtran}
\bibliography{mybib}
\end{document}